# Thermoelectric properties of metal-semiconductor percolation composites

Oleksandr Porubaniyi[1], Sergij Tomko[1], Myhailo Cherkez[1], Viktor Podolskiy,[2,*] and Radion Cherkez[1]

[1)] Department of Thermoelectricity and Medical Physics, Yuriy Fedkovych Chernivtsi National University, 2 Kotsiubynsky str., 58012,Ukraine

[2)] Department of Physics and Applied Physics, University of Massachusetts Lowell, Lowell, MA, 01854, USA
[*]corresponding author: viktor_podolskiy@uml.edu

We analyze the thermoelectric response of composite materials comprising homogeneous semiconductor hosts with randomly distributed metallic inclusions. We demonstrate that the properties of such materials are strongly affected by the percolation transition (with background conductivity). We show that in the vicinity of the percolation threshold, the power factor and the thermoelectric figure of merit strongly depend on both the concentration and arrangement of inclusions. Finally, we identify promising regimes for enhancing the power factor in realistic composites.

## I. Introduction

Thermoelectricity and thermoelectric cooling are crucial in space exploration, industrial waste heat recovery, and in compact refrigeration for medical, electronic, and military devices[1,2]. The quest to find the perfect thermoelectric material has resulted in multiple candidates, including various composite media, structurally inhomogeneous materials that combine the properties of multiple components[2-20]. While the majority of these materials are all-semiconductor heterostructures, a broad class of metal/semiconductor composites recently re-emerged as promising candidates for the future of thermoelectric materials[12-14,17,18].

In particular, Bergman and coauthors [15,16] considered analytically the perspectives of combining high-performance thermoelectric semiconductors with noble metal inclusions, predicting that such composites

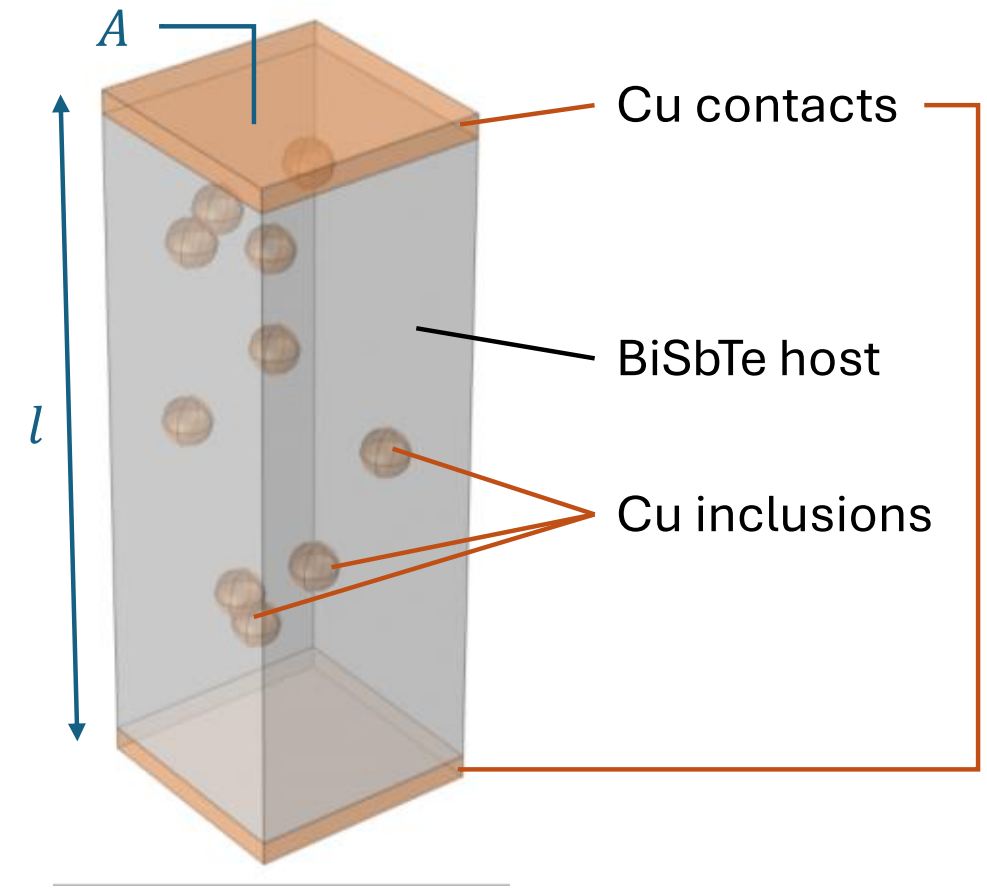

Fig.1. Schematic of the thermoelectric composite analyzed in this work (10 spherical inclusions are shown; hundreds of inclusions are used in the study), along with geometric parameters used in the model

can achieve significant improvement in the power factor at a minor cost of thermoelectric figure of merit zT. As with any analytical model, the results of these studies are limited to composites whose highly conductive inclusions occupy relatively small volume and have relatively simple shapes (spheres/ wires/ layers, and other ellipsoids) [15-23]. However, while realistic composite materials have complex internal structure, existing research efforts have mostly been focused on multilayered materials or on low-metal-concentration (Maxwell-Garnett) limit of two- and three-dimensional materials. At the same time, it is known that the properties of complex media may significantly deviate from the predictions of simplified medium theories even in the limit of small inclusion concentration [24-29]. In fact, the problem of understanding the properties of homogenizable composites in electrostatic limit has only been very recently solved [24]. Here we use rigorous numerical analysis of realistic three-dimensional composites to address the long-standing issue of performance of complex thermoelectric/metal composites. We demonstrate that the power factor and the figure of merit of the thermoelectric composite strongly depend on the metal concentration. Moreover, for a range of metal concentrations, the properties of the composite strongly depend on the arrangement of inclusions. Our analysis demonstrates a wide parameter space where the power factor of the composite can be significantly improved at a relatively minor cost of the thermoelectric figure of merit.

## II. Methods

The geometry of the structures analyzed in this work is shown in Fig.1. Similar to previous works, we consider a bulk semiconductor host (BiSbTe in our analysis) with metallic (Cu) inclusions. For simplicity, we use spherical inclusions. However, in contrast to previous analytical and numerical studies, we position the 100-μm-radius inclusions randomly and allow these inclusions to partially overlap, potentially forming extended clusters. The model comprises a region of the composite of thickness $l = 3mm$ with a unit cell area of $A = 10^{-6}m^2$. To avoid numerical and meshing instabilities, the spheres are placed on a square grid with the size of 25 μm and are not allowed to touch the boundaries of the unit cell, with minimum distance between the cell boundaries and spheres being 5 μm (we discuss the implications of such design and possible artifacts in text below). Periodic boundary conditions are applied along the in-plane directions to simulate response of the spatially-infinite material. The composites analyzed comprise 10…300 spheres per composite. To assess the finite-size effects that are common to percolation composites, we have also evaluated properties of cells that are 1.5x larger in every direction ($l = 4.5mm,\ A = 2.25\ 10^{-6}m^2$), comprising 300…700 spheres each.

The DC properties of a random mixture with metallic inclusions are known to exhibit a dielectric/metal phase transition. The volume fraction corresponding to this transition is known as the percolation threshold $p_c$ [25-32]. In the vicinity of the percolation threshold, local distribution of electric field, as well as distribution of electric currents are extremely inhomogeneous and the averaged conductivity follows scaling behavior $(\sigma - \sigma_b) \propto (p - p_c)^t$ for $p > p_c$ and $(\sigma - \sigma_b) \propto (p_c - p)^{-s}$ for $p < p_c$. Critical exponents $t, s$ depend on the dimensionality of the network, and – in case of the network with non-zero background conductivity $\sigma_b$ – on $\sigma_b$[27-31]. In general, the behavior of percolation networks of finite size may somewhat deviate from these universal laws due to finite-size effects [30]. One of the goals of the current study is to assess the implications of percolation transition for thermoelectric properties of the composite.

Thermoelectricity represents the interplay between the electric current density $\vec{j}$ and heat fluxes $\vec{J_q}$ in the system. Explicitly,

$$\vec{j} = -\sigma\,(\vec{\nabla}\phi + S\vec{\nabla}T) \qquad (1)$$

$$\vec{J_q} = TS\,\vec{j} - \lambda\,\vec{\nabla}T$$

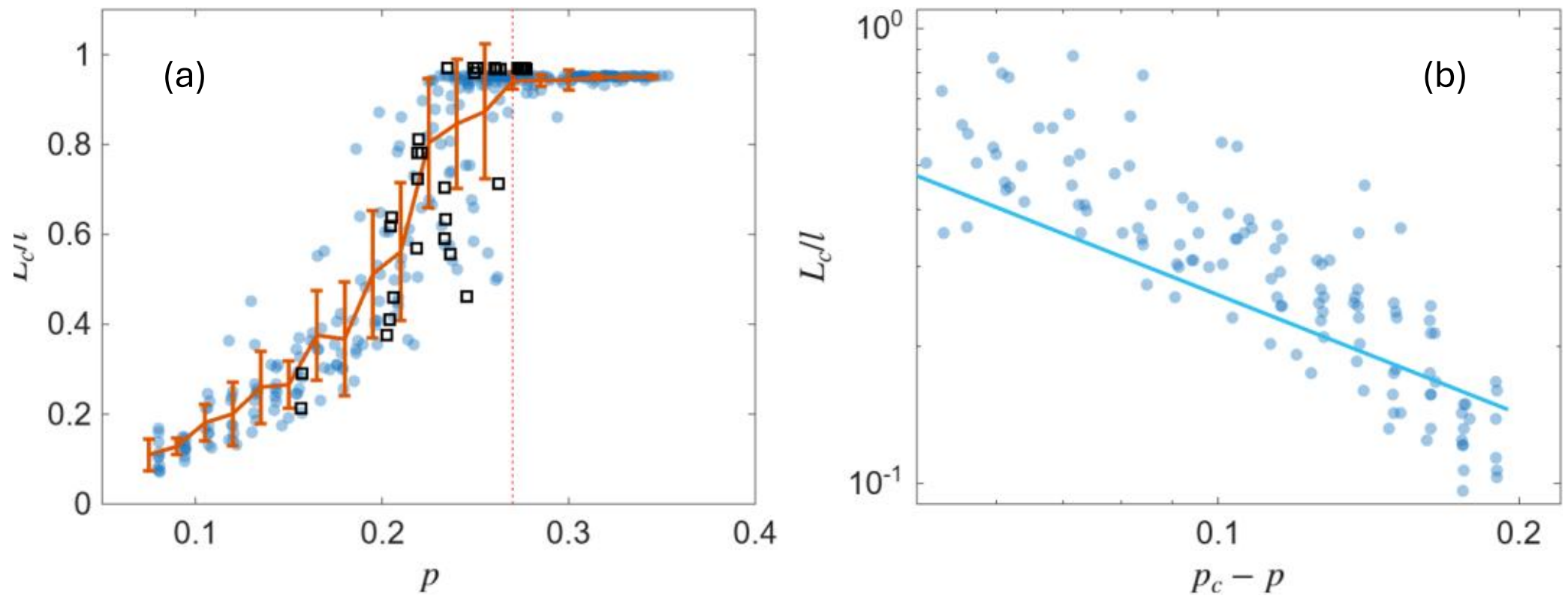

Fig.2. (a) Length of the largest conductive cluster (normalized to the height of the unit cell $l$) as a function of metal concentration $p$. Each blue dot represents one realization of the composite; orange line represents mean and standard deviation of the distribution. Black squares represent realization of networks in large (linear dimensions x1.5) unit cells. Red dotted line corresponds to percolation threshold. (b) scaling properties of the maximum cluster size below percolation; solid line corresponds to $\propto (p_c - p)^{-\nu}$

with $\sigma$ and $\lambda$ being electrical and thermal conductivities, and $S$ being Seebeck coefficient that describes the thermoelectric coupling, with $\phi$ and $T$ being electrostatic potential and temperature[1].

Here we use the commercial finite-element partial differential equation (PDE) solver COMSOL Multiphysics[33] to analyze electric and thermal fluxes of the composites, in conjunction with the interaction between the two. Three separate simulations are performed. In the first two runs, we neglect the (weak) thermoelectric coupling to calculate the effective electrical and thermal conductivities as $\sigma = Il/VA$ and $\lambda = ql/\Delta TA$, with $I, V, q, \Delta T$ being DC current, applied voltage, thermal flux, and temperature difference across the composite, respectively. The last simulation provides self-consistent solution of thermoelectric generation at the fixed temperature difference $\Delta T = 10^o C$ between the top and bottom surfaces, with baseline temperature of the composite being $0^o C$. The results of this last solution yield the effective Seebeck coefficient of the composite, calculated as $S = \Delta V / \Delta T$, with $\Delta V$ being the thermoelectrically induced potential difference across the composite. Multiple (over 700) composites with randomly distributed conducting spheres are generated and their properties are analyzed; in addition ~30 composites with large unit cells and concentrations close to percolation threshold are analyzed.

The model uses temperature-dependent electrical and thermal conductivities of materials. In the regime used in this work, conductivities of the constituents are given by $\lambda_{Cu} \simeq 400\ W/m \cdot K$, $\lambda_{BiSbTe} \simeq 1.28\ W/m \cdot K$, $\sigma_{Cu} \simeq 6 \cdot 10^7 S/m$, and $\sigma_{BiSbTe} \simeq 6 \cdot 10^4\ S/m$.

The performance of the composite is evaluated using two standard thermoelectricity metrics, power factor $P_f = S^2\sigma$ that characterizes the electrical power output of the device at a given temperature gradient and the dimensionless figure of merit $zT = S^2\sigma T/\lambda$ that characterizes the overall thermoelectric efficiency [1,2,34,35]. Both of these metrics are calculated using the effective composite parameters, as defined above; $T = 273\ K$ is used for $zT$ calculations.

## III. Results

As the composite is assembled from individual spheres, some of the spheres touch or overlap, forming randomly-shaped and -sized clusters. The dependence of the length of the largest cluster (normalized by the height of the unit cell) as a function of concentration of the composite is illustrated in Fig.2. It is seen that the size of such a cluster rapidly increases with concentration, reaching the size of the unit cell when $p \simeq 27\%$. This concentration determines the value of percolation threshold in our composite. Notably, our analysis is not significantly affected by the finite-size effects since the dependence of the normalized cluster length on metal concentration does not depend on the size of a unit cell.

According to percolation theory, the size of the largest cluster in the network is expected to scale as $L_c \propto (p_c - p)^{-\nu}$ with critical exponent $\nu \simeq 0.87$ [27]. This expected behavior is shown in Fig. 2(b).

Fig.3 illustrates the dependence of electrical and thermal conductivities of individual composites on concentration of metal (Cu) inclusions – the only parameter that controls the response of composites in the effective medium regime. As mentioned above, in our model inclusions are not allowed to touch the highly conductive Cu electrodes. The (thin) BiSbTe layer between the electrodes and the Cu inclusions plays the role of the low-conductivity layer, connected in series to the bulk of the thermoelectric composite. We therefore subtract the contribution of these low-conductance layers (as well as contributions of the contacts themselves) from the conductivity derived from FEM solutions in the analysis reported in Fig. 3.

Dots in Fig.3 represent the data for individual realizations of random composites. Note that both electrical and thermal conductivities follow similar behavior. Three regions can be clearly identified: when the concentration changes from 0 to $\sim 18\%$ conductivities increase almost linearly with concentration, growing by approximately a factor of two. This behavior is consistent with the effective medium description derived in Refs.[15,16,21-23]:

$$k_{EMT} = k_{BiSbTe} \frac{1+2\,p\frac{k_{Cu}-k_{BiSbTe}}{k_{Cu}+2\,k_{BiSbTe}}}{1-p\frac{k_{Cu}-k_{BiSbTe}}{k_{Cu}+2\,k_{BiSbTe}}}, \qquad (2)$$

with $k$ standing for electrical ($\sigma$) or thermal ($\lambda$) conductivities, respectively, and with subscript EMT standing for effective medium parameters. Effective medium predictions are shown with dashed lines in Fig. 3(a,b)

In the second region, when concentration changes from ~18 to ~27%, conductivities rapidly grow, changing by almost an order of magnitude. Notably, this growth of average conductivity is accompanied by strong variation between conductivities of individual realizations of random materials. This region is not described by effective medium response since the parameters of the structure strongly depend on the arrangement of the inclusions[29]. Finally, when metal concentration crosses $p_c \simeq 27\%$, there is a noticeable change of slope in the dependence of conductivities on the volume fraction of the inclusions.

To systematically analyze the properties of the random composites we sort the data by the volumetric concentration of metal and group the composites into bins that have concentration “width” $\Delta p = 1.5\%$. We then calculate the average value and the standard deviation of each quantity of interest within individual bins. The dependence of mean values of conductivities on the average metal concentration within the composite is shown as solid lines in Fig.3(a,b). In agreement with the above discussion, the properties of composites in the effective medium regime are almost independent of the exact arrangement of inclusions (the variance of the resulting parameters is very small); with the onset of percolation transition, the variance of the parameters increases dramatically. Once again, it is seen that our analysis is not significantly affected by the finite-size effects since the behavior of larger composites is in line with the smaller-composite counterparts.

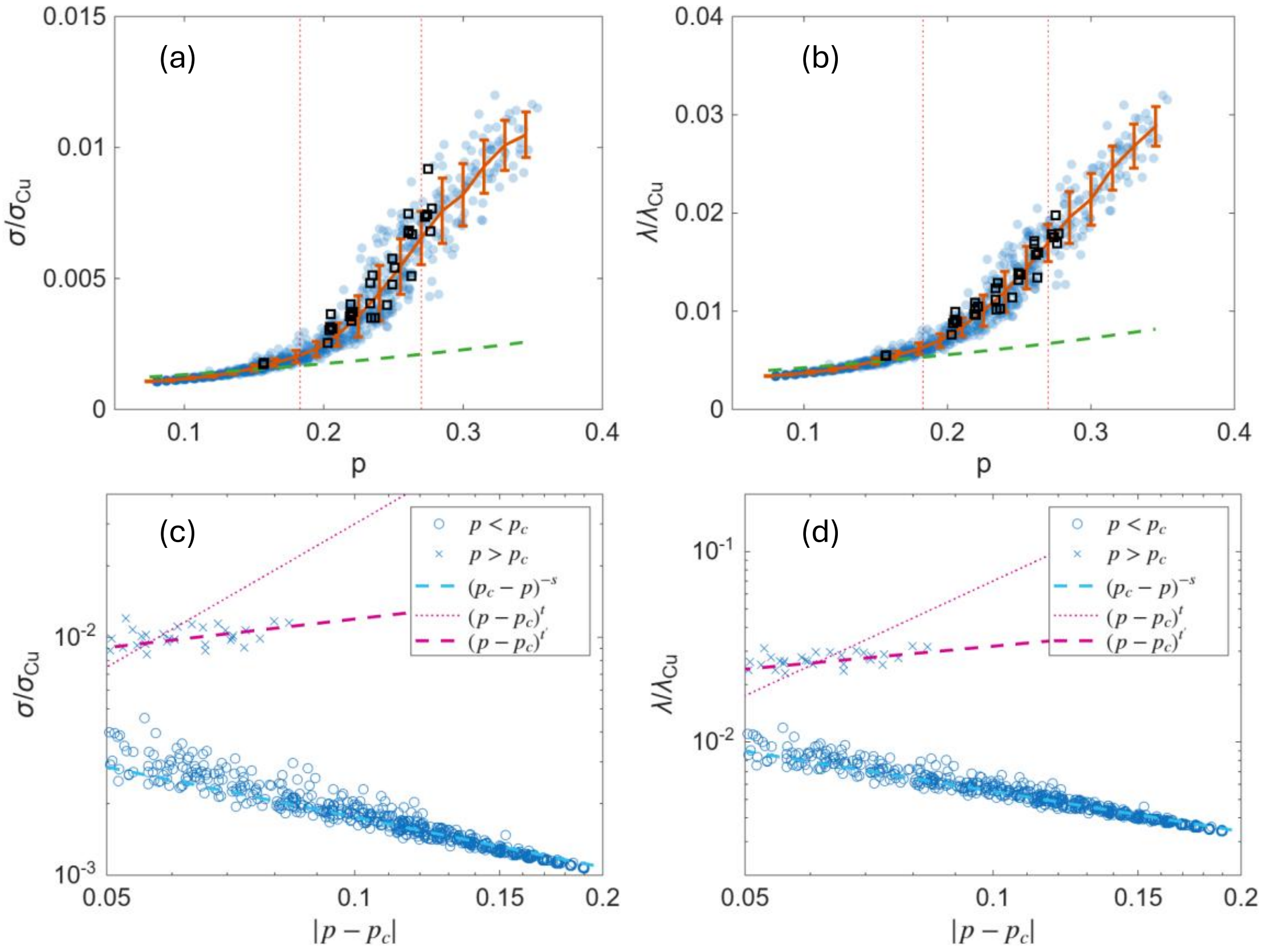

Fig.3 Electrical (a,c) and thermal (b,d) conductivities of composites as a function of volumetric metal concentration; each symbol represents one realization of a composite in nominal (circles) and large (squares) unit cells; solid lines illustrate bin-averaged properties of composites with nominal unit cells; dotted lines in (a,b) represent the limit of separate effective medium regime and percolation threshold; dashed lines represent Eq.(2); (c,d) scaling properties of conductivities; lines represent scaling fits (see text for details);

Fig.3(c,d) illustrates the scaling behavior of electrical and thermal conductivities below and above $p_c$. It is seen that the composites follow the expected power-law dependence, with well-established critical exponents $s \simeq 0.7, t \simeq 2$ [27] reasonably describing the behavior of electrical conductivity and of thermal conductivity below $p_c$. At the same time, the behavior of conductivity for $p > p_c$ deviates from the predictions of conventional percolation theory [31]. Our analysis suggests that renormalized exponent $t' \simeq 0.4$ adequately describes the response of the composite.

Fig.4 illustrates our analysis of thermoelectric performance of the composites in our studies. The effective Seebeck coefficient is almost independent of metal concentration in the effective medium regime, decaying by $\sim 40\%$ for higher concentrations. Interestingly, the Seebeck coefficient increases for $p > p_c$ in composites with our nominal unit cell. At the same time, the analysis of structures with larger unit cells reveals that this increase is probably a manifestation of finite-size effects in our model. The spheres (that do not touch the sides of the unit cell) form conducting clusters that are "suspended" within the semiconductor matrix. In the limit of relatively high metal concentration, the composite simulated resembles an array of such clusters, vertically aligned with each other, with each cluster fully encompassed within semiconductor. Larger unit cells yield stronger heterogeneity of individual clusters, bringing the dependence of the Seebeck coefficient in line with expectations – $S$ monotonically decreases with increase of $p$.

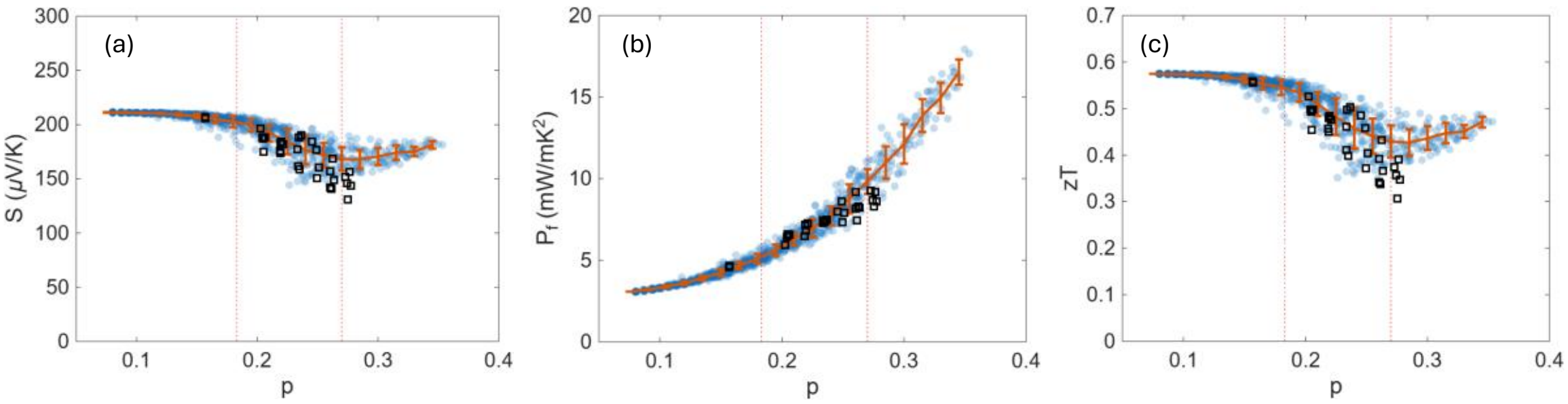

Fig.4. Dependence of the effective Seebeck coefficient S (a), power factor $P_f$ (b) and thermoelectric figure of merit $zT$ (c) of the composites on the volume fraction of Cu, $p$; symbol represents a single composite with nominal (circles) or large (squares) unit cell; orange lines represent statistical analysis as discussed in the text; dotted lines show the limit of effective medium response and percolation threshold

The above numerical artifact, however, points to a possible new degree of freedom in the optimization of thermoelectric composites: forming materials with pre-designed periodic internal structures, in some analogy to recent successes in optical and acoustic metamaterials [36,37]

The power factor $P_f = S^2\sigma$, dominated in our case by the electric conductivity, grows with metal concentration, increasing by a factor of two over the effective medium region, with an extra factor of two from percolation threshold until $p \sim 30\%$. The power factor (through its dependence on $S$) is somewhat affected by the finite-size effects described above.

Finally, the thermoelectric figure of merit $zT = P_f T/\lambda$, largely follows the behavior of Seebeck coefficient, but with more pronounced dependence on metal concentration. zT is virtually unchanged over the effective medium region, followed by a decay by $\sim 40\%$ over the range $p \in [0.18 \ldots 0.3]$ when significant fraction of the current flows over extended metallic clusters. In our analysis, the behavior of $zT$ for larger metal concentrations strongly depends on unit cell size.

## IV. Discussion

Compared with previous studies focused on effective medium response[15-20] and on layer-based materials[12-14] we focus on thermoelectricity in percolation-driven composites, with emphasis on metal concentrations $p \simeq p_c$ and $p > p_c$. Our analysis clearly shows that metal cluster formation within the composite makes the materials very sensitive to the presence of metal: while layer-based analysis[12] suggested that $zT$ would be independent of metal over the broad range ($p \in [0,90\%]$), our analysis suggests significant reduction of the figure of merit for relatively small metal concentrations (p~25%), in agreement with previous studies of two-dimensional random resistor networks [29].

Given the importance of the figure of merit $zT$ for thermoelectric applications, the best-possible performance is expected to be close to percolation threshold, where the composite yields doubling of the power factor with virtually no reduction of $zT$, as compared with the bare semiconductor. Notably, in this regime the composite exhibits minimal dependence of the effective properties on arrangements of individual inclusions. The performance of the random composites can be further enhanced if, as a result of embedding metal clusters into the semiconductor host, thermal conductivity of the host decays, as suggested in Ref.[12] and observed in experiments in [17,19,20]. Such a decay will be accompanied by the increase of the figure of merit ($zT$), with no expected effect on the power factor.

Our analysis also suggests that the behavior of composites can be improved (with significant increase in $P_f$ and zT potentially reaching the value of the host semiconductor) in composites that comprise aligned but separated conductive clusters.

## V. Conclusion

To conclude, we have analyzed the behavior of thermoelectric composites that comprise a homogeneous semiconductor host with randomly distributed metallic inclusions. While our results are qualitatively consistent with previous studies (increase of metal concentration yields a significant increase in power factor and a small reduction in the figure of merit), the quantitative response of realistic three-dimensional materials significantly deviates from predictions of effective medium response and from the properties of the layered media. Based on the universality of Eq.(1), we expect that our results (the breakdown of the effective medium model, the effect of percolation transition on conductivities, power factor, and thermoelectric figure of merit), illustrated on the example of BiSbTe-Cu composite, are applicable to any combination of materials where the semiconductor host is permeated with inclusions of much higher electrical and thermal conductivities. Inclusion shape plays a crucial role in defining the percolation threshold with more elongated inclusions leading to the smaller values of $p_c$ with ellipsoid-based composites having lower percolation threshold than their sphere-based counterparts [38]. Such a reduction of percolation threshold associated with elongated conductive components suggests that spherical inclusions may offer advantages for optimizing the power factor of composites with random inclusions placement and orientation. Our study provides an outlook for a new generation of thermoelectric composite materials.

## Author Declarations

The authors have no conflicts to disclose

## Data Availability Statement

All relevant data is provided in this work. The FEM model developed in this study is available online [39]. Other data that supports the findings of this study is available from the corresponding author upon reasonable request